\documentclass[12pt]{article}
\usepackage[english]{babel}
\usepackage[utf8x]{inputenc}
\usepackage{amsmath}
\usepackage{amssymb}
\usepackage{mathtools}
\usepackage{graphicx}
\usepackage{multicol}
\usepackage[margin=1in]{geometry}
\usepackage{times}
\usepackage{setspace}

\usepackage{amsthm}
\usepackage{mathtools}

\theoremstyle{definition}

\title{Curricular Complexity Versus Quality of Computer Science Programs}
\author{Gregory L. Heileman$^{\dagger\natural}$ \and Hayden W. Free$^\ddagger$ \and Johnny Flynn$^\ddagger$ \and Camden Mackowiak$^\ddagger$  \and Jerzy W. Jaromczyk$^\ddagger$ \and Chaouki T. Abdallah$^\star$ \\ \\
  \normalsize $^\dagger$heileman@arizona.edu \\
  \normalsize Department of Electrical \& Computer Engineering \\
  \normalsize University of Arizona  \\ \\
  \normalsize $^\ddagger$\{hayden.free, john.flynn, camden.mackowiak, jurek\}@uky.edu \\
  \normalsize Department of Computer Science \\
  \normalsize University of Kentucky \\ \\
  \normalsize $^\star$ctabdallah@gatech.edu \\
  \normalsize Department of Electrical \& Computer Engineering \\
  \normalsize Georgia Institute of Technology \\ \\
  \normalsize  $^\natural$corresponding author }

\date{} 

\begin{document}

\maketitle

\section{Abstract}
In this research paper we describe a study that involves measuring the complexities of undergraduate curricula offered by computer science departments, and then comparing them to the quality of these departments, where quality is determined by a metric-based ranking system.  The study objective was to determine whether or not a relationship exists between the quality of computer science departments and the complexity of the curricula they offer.  The relationship between curricular complexity and program quality was previously investigated for the case of undergraduate electrical engineering programs, with surprising results.  It was found that if the US News \& World Report Best Undergraduate Programs ranking is used as a proxy for quality, then a statistically significant difference in curricular complexities exists between higher and lower quality electrical engineering programs.  Furthermore, it was found that higher quality electrical engineering programs tend to have lower complexity curricula, and vice versa.  In the study reported in this paper, a sufficient amount of data was collected in order to determine that an inverse relationship between program quality and curricular complexity also exists in undergraduate computer science departments.  This brings up an interesting question regarding the extent to which this phenomenon exists across the spectrum of STEM disciplines. 
   
\section{Introduction}\label{intro}
Various metrics can be used to quantify the complexity of the curricula associated with academic programs. One useful representation of curricular complexity involves decomposing it into two independent parts, one that involves structural factors, i.e.,  \emph{structural complexity}, and the other that involves instructional factors, i.e., \emph{instructional complexity}.  In this study we used a metric based solely on curricular structure that is obtained by representing a curriculum as a graph, were the vertices are the required courses in a curriculum, and the edges represent the pre- and co-requisite relationships between these courses.  Instructional complexity, on the other hand, involves factors such as the quality of teaching, the support services provided to students, the inherent difficulty of course concepts, etc.  It can be shown that if instructional complexity is held constant, then structural complexity directly relates to the difficulty of progressing through the courses in a curriculum.  More specifically, for two programs with similar instructional complexity, and serving similar student populations, the one with lower structural complexity will have higher graduation rates~\cite{HeAbSlHi:18}. Thus, from the perspective of student success, structural complexity matters. Furthermore, according to this complexity metric, STEM programs tend to be among the most complex curricula at a university; this is attributed to the large number of prerequisites that accompany many of the courses in STEM programs, as well as the long prerequisite chains that tend to exist in these curricula. 

\begin{figure*}
 \vspace*{-0.2in}
 \centerline{\includegraphics[width=4.75in]{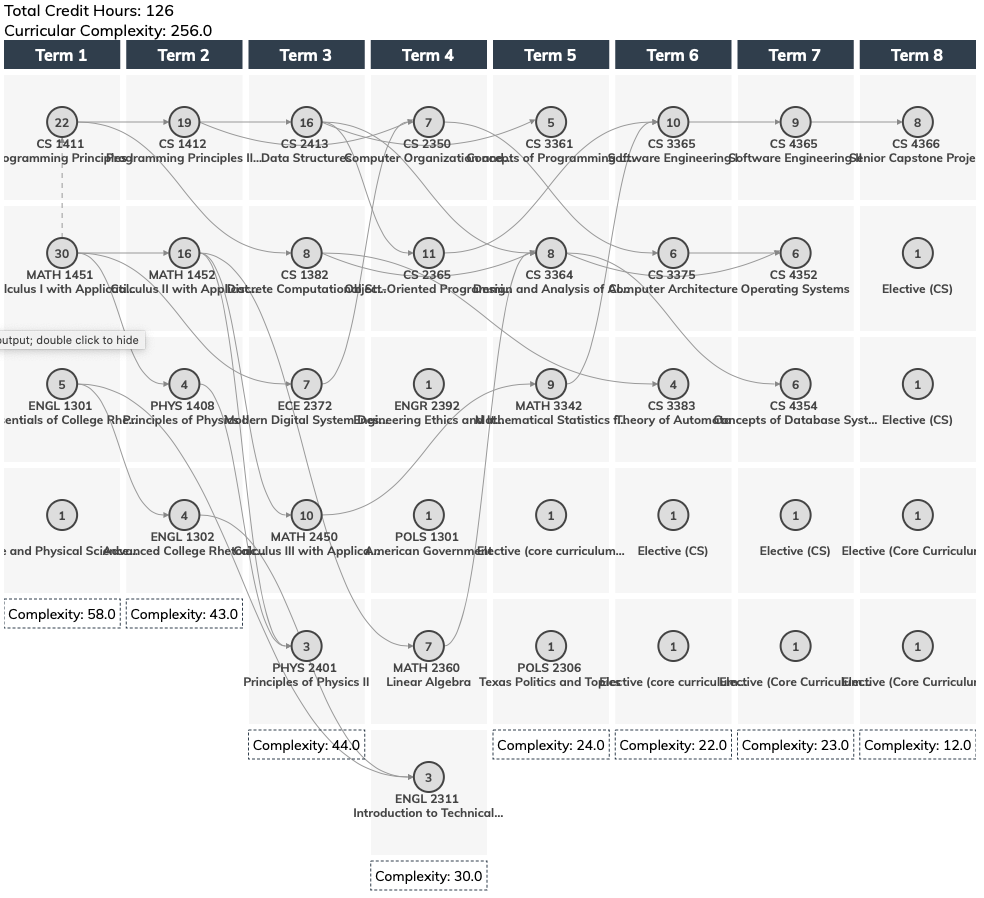}}
 \begin{center}
 {\bf\small (a)}
\end{center}
 \vspace*{0.2in}
 
 \centerline{\includegraphics[width=4.75in]{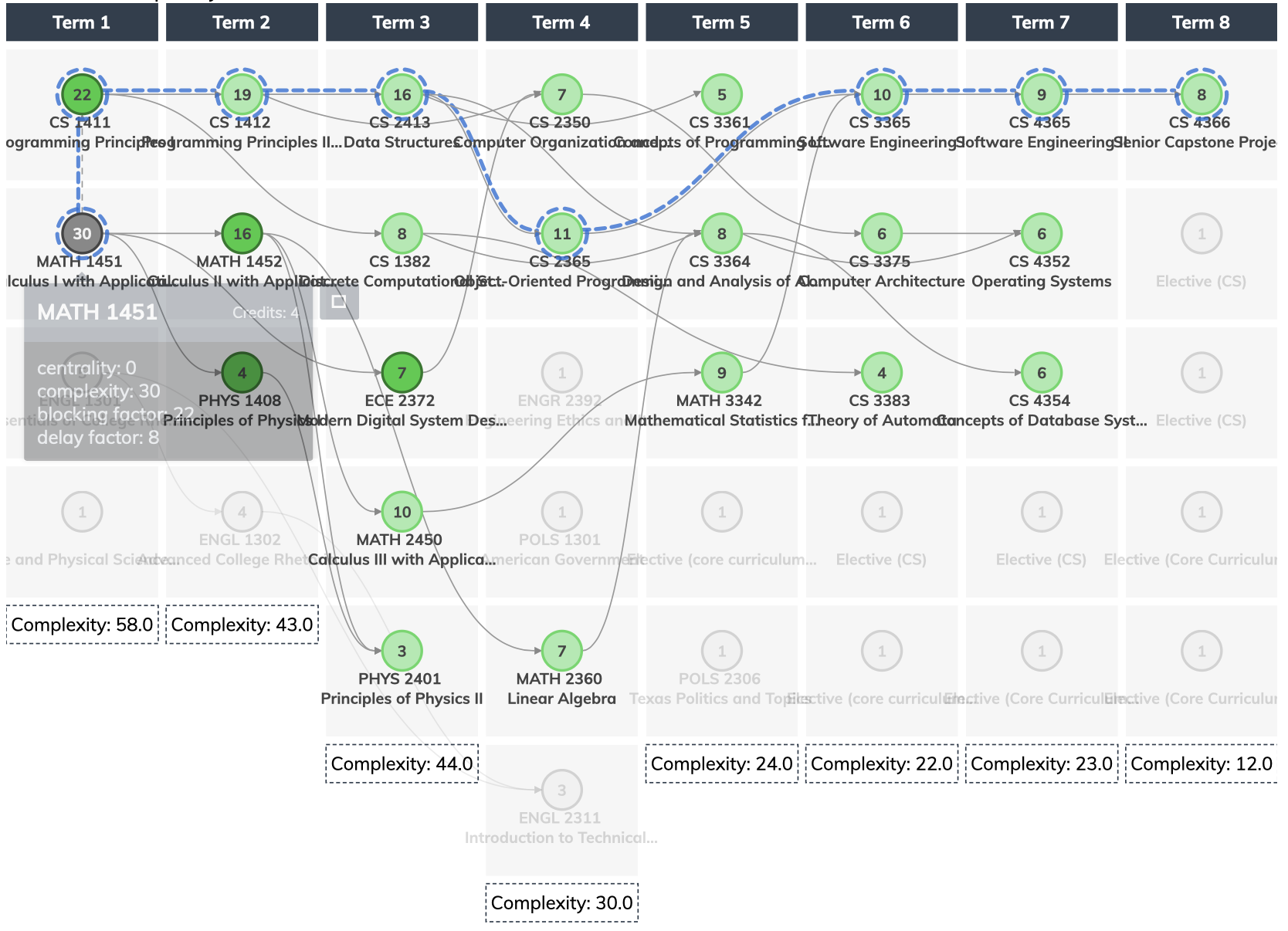}}
  \begin{center}
  {\bf\small (b)}
 \end{center}
 \caption{{\bf (a)} The 2018-19 degree plan for a computer science program at a school in the southwest United States. {\bf (b)} Highlighting  Calculus I (gray) in the first term, there are 22 courses blocked by Calculus I (green), and the longest path in this curriculum (blue-dashed line) contains 8 courses.}\label{DP1}
\end{figure*}

In order to understand the structural curricular complexity measure, consider the computer science curriculum shown in Figure~\ref{DP1}. 
The curriculum shown in Figure~\ref{DP1}~(a) is organized as a 4-year (8-term) degree plan.  The structural complexity measures for each course are shown inside the course vertices in this figure. The total structural complexity of the courses associated with each term are shown at the bottom of each term, and the overall structural complexity of the entire curriculum, given as the sum of the term complexities, equals 256.  The structural complexity measure for a single course is computed as a linear combination of the delay and blocking factors, which are demonstrated in Figure~\ref{DP1}~(b). Specifically, in this figure, the Calculus~I course is highlighted in dark gray.  The number of courses in the curriculum for which Calculus I serves as a prerequisite, or as the prerequisite of a prerequisite, etc., are shown in green in this figure.  That is, the green courses are those that rely upon the ability to first pass Calculus I.  For this reason the inability to pass Calculus~I is said to block 22 other courses in this curriculum, and therefore Calculus~I has a blocking factor of 22.  

The delay factor of a course, on the other hand, is determined by the length of the longest path that a course is on. Long paths in a curriculum are important---if a students does not pass a course on a long path, they are delayed in moving to any other course on that path.  Furthermore, the number of terms spanned by a longest path in a curriculum provides a lower bound on the number of terms required to complete that curriculum. For instance, in Figure~\ref{DP1}~(b), the delay factor associated with Calculus~I is 8 because there are eight courses on the longest path containing Calculus~I. 

\begin{figure*}
 \centerline{\includegraphics[width=4.75in]{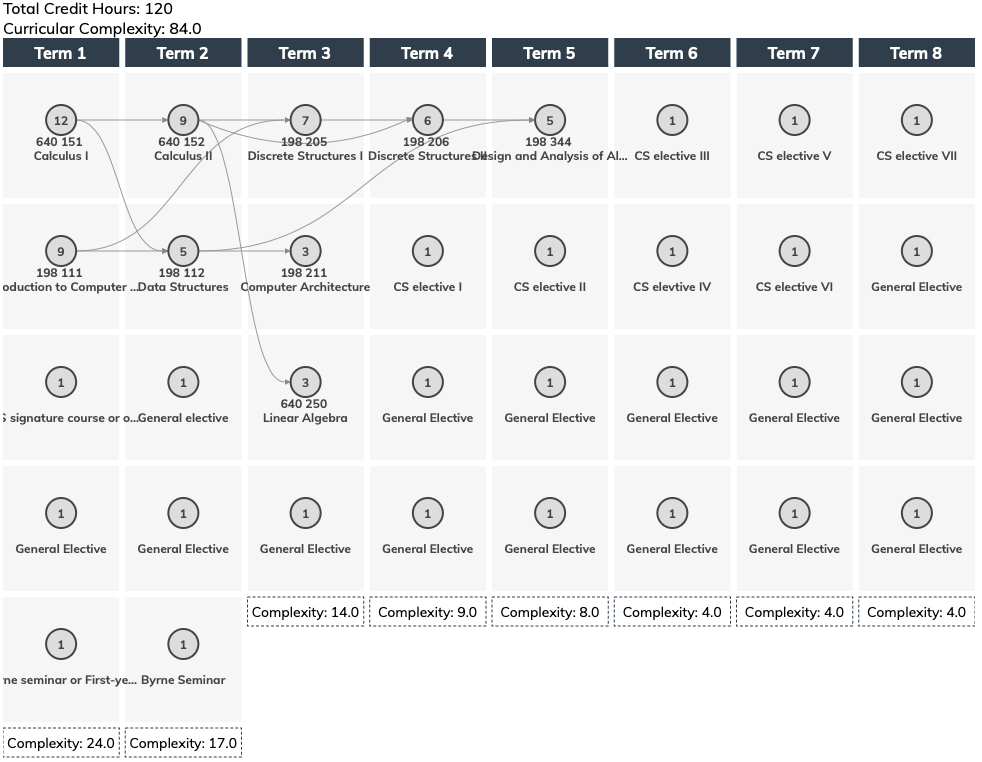}}
 \begin{center}
 {\bf\small (a)}
\end{center}
 \vspace*{0.2in}
 
 \centerline{\includegraphics[width=4.75in]{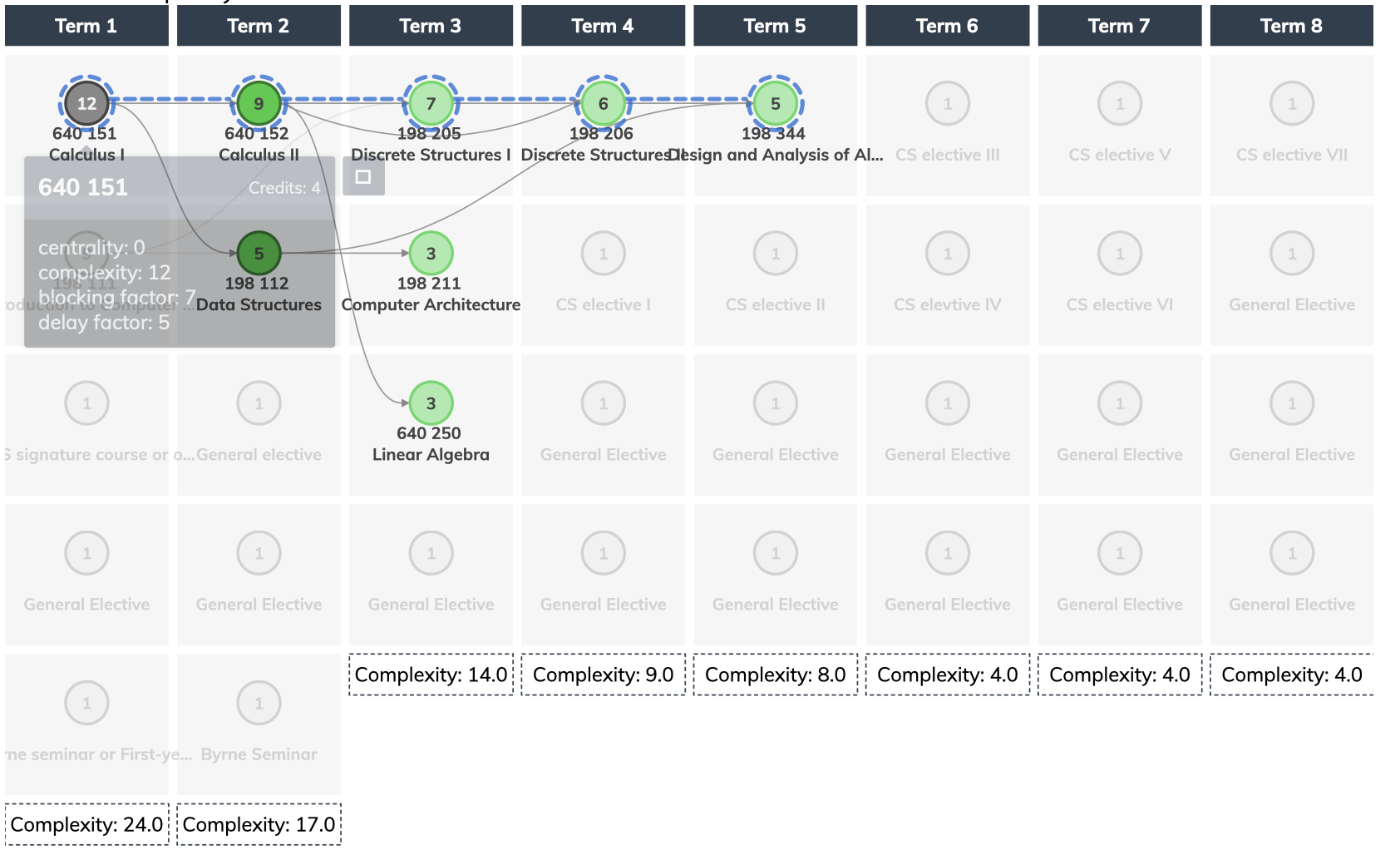}}
  \begin{center}
  {\bf\small (b)}
 \end{center}
 \caption{{\bf (a)} The 2018-19 degree plan for a computer science program at a school in the northeast United States. {\bf (b)} Highlighting  Calculus I (gray) in the first term, there are 7 courses blocked by Calculus I (green), and the longest path in this curricula (blue-dashed line) contains 5 courses.}\label{DP2}
\end{figure*}

In order to get a sense of the variability that exists among the curricula of different computer science programs, consider the curriculum shown in Figure~\ref{DP2}, which has a structural complexity of 84.
The overall complexity of the curriculum in Figure~\ref{DP1} is more than three times as complex as the one shown in Figure~\ref{DP2}. The question, then, is does this extra curricular complexity generally correspond to a higher quality output?  Because it is not feasible to assess the extent to which learning outcomes are attained by the students in all of the various undergraduate computer science programs, we will instead let the overall ranking of a computer science department serve as a proxy for the average quality of the students who graduate from the programs in that department. Further rationale for this approach is provided in the Methodology section below.

An interesting feature of this study is that many of the programs considered have the same ABET (CSAB) accreditation. Thus, they share the same set of student learning outcomes, and are therefore equivalent in the sense that they all produce competent computer scientists. Furthermore, in all of the programs we considered, the courses that are offered appeared very similar across the curricula. For instance, nearly all include Calculus I and Programming I in the first term of the freshman year, a Discrete Math course in the sophomore year, an Algorithms course in the junior year, etc.  There is, however, extreme variability in the way different computer science programs structure their curricula. Some programs include a larger number of courses (and credit hours) and are tightly prescribed in that they stipulate a larger number of prerequisites for key courses.  Other programs have fewer courses (e.g., they meet the 120-credit-hour minimum that regional accreditors expect) and they provide more freedom by having fewer prerequisites for key courses, and a proportionally larger number of elective courses.

\section{Methodology}\label{methodology}
In this study we used the rankings provided by the CSRankings of computer science departments~(restricted to universities in the United States) as a proxy for program quality.  
CSRankings are entirely based upon metrics related to extent to which a department's faculty participate in prestigious publication venues~\cite{CSRankings}.
That is, for the purpose of this study, we assume the highest ranked computer science departments are synonymous with the highest quality computer science undergraduate  programs.  We acknowledge the concerns that are routinely expressed concerning rankings such as these~\cite{By:18,Gl:11}.  However, the use rankings as a surrogate for brand value, and the notion that brand value is elevated through the accumulation of institutional quality, has been previously established~\cite{Lo:13,RoLoCr:18}.
In addition, it should be noted that this study uses aggregations of departments within tiers, and the statistics associated with these aggregations.  Thus, the specific rankings of the departments within the  tiers are irrelevant, all that matters is the tier in which a department is placed. Upon inspection of the departments within each tier, we believe that knowledgable and  impartial observes would agree that the three tiers constructed in this study are highly correlated with program quality. Finally, it is important to recognize that this study was constructed to demonstrate whether or not there is a statistically significant \emph{correlation} between program complexity and program quality.  We do not prove the existence of any \emph{causal factors} that may have led to this correlation; however, we discuss possible causal relationships in the Discussion section below.

\section{Experiment Design}\label{design}
The question of interest in this study is whether or not curricular complexity is related to program quality. In order to answer this question, we constructed 
an analysis of variance~(ANOVA) experiment that involved partitioning the departments in the CSRanking list according to their decile within the ranking. From 
these, three groups were created as follows. A top tier of departments defined as those in the first decile of the ranking. A middle tier, defined as the set of departments 
in the fifth and sixth deciles of the ranking that are equidistant from first to last ranked departments. A bottom tier of departments comprised of the departments ranked in 
the bottom two deciles. The null hypothesis is:
\begin{quote}
 \emph{There is no difference between the mean values of the structural curricular complexities of those computer science programs associated with the top, middle and bottom tier departments.}
 \end{quote}
The alternative hypothesis is then:
\begin{quote}
 \emph{At least one of the structural curricular complexity mean values of a department tier differs significantly from the mean values of the other two tiers.}
\end{quote}

\begin{figure}
 \centerline{\includegraphics[width=4in]{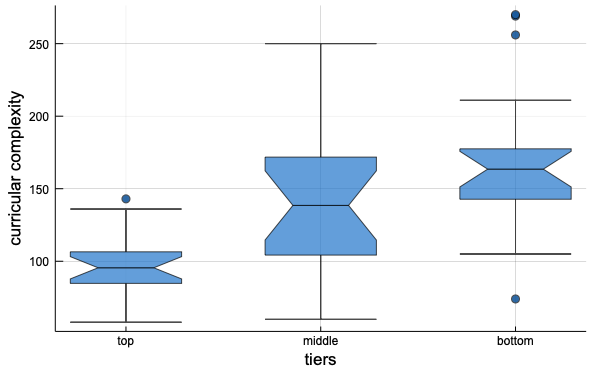}}
 \caption{Box-and-whisker diagram constructed from samples taken from each of the three tiers. Each box encompasses the upper~(75\%)  and lower~(25\%) quartiles, i.e., the interquartile range~($IQR$), of the curricular complexities of the curricula in the sample, the line inside the notched box is the median value of the sample, and the whiskers show the extreme curricular complexity scores (excluding outliers) within each sample. Outliers (a curricular complexity score greater/less than $1.5$ times the upper/lower quartile) are shown as dots.}\label{box-whisker}
\end{figure}

A notched box-and-whisker diagram constructed using samples taken from each of the three ties is provided in Figure~\ref{box-whisker}.
Notice that the median values for these samples are quite different from one another, and that the smallest structural complexity variability appears in the sample from the top tier of schools.  The notches in the boxes represent approximate confidence intervals about the median values.  They are constructed using:
\[
  m_i \pm 1.57 \times IQR/\sqrt{n_i},
\] 
where $m_i$ is median value, $IQR_i$ is the interquartile range, and $n_i$ is the number of observations in the $i$-th sample~\cite{ChClKlTu:83}.  The fact that some of these notches do not overlap along the complexity axis provides some evidence that the null hypothesis may not be true.  Hence, a more extensive ANOVA analysis is warranted. 

\section{ANOVA Analysis}
The ANOVA analysis in this study involves random sampling of departments within each of the three tiers defined above. In order to ensure the analysis is able to distinguish between 
actual structural curricular complexity differences among the tiers, rather than random variation, a sufficient number of random samples must be collected. Under the assumption that the structural complexity distributions within the tiers are approximately normal, with variance $\sigma^2$, the number of samples that should be randomly selected from each tier is given by
\begin{equation}
 n = \left({\sigma Z \over E}\right)^2,
\end{equation}
where $Z$ is the confidence interval expressed using deviation within the standard normal distribution, and $E$ is the margin of error. To obtain an 
estimate of $\sigma$, pilot samples from each of the three tiers were taken, yielding the estimate $\hat{\sigma} = 60$. For a 95\% confidence interval, 
which corresponds to $Z = 1.96$, the margin of error will be 30 structural complexity points, i.e., 15 points on either side of the mean for a tier. Using 
these values in the equation provided above leads to sample sizes of $n_1 = n_2 = n_3 = 20$, where $n_1, n_2$ and $n_3$ are the sample sizes for 
the top, medium and bottom tiers, respectively. Thus, by sampling at least 20 departments from each tier, we can have 95\% confidence that the error in 
this analysis will be by no more than 30 structural complexity points.

According to the sample size analysis provided above, 20 departments were randomly sampled from each tier.  From the top decile of the CSRankings, the following departments were randomly selected, and the undergraduate computer science curricula were extracted from information contained on their departmental websites: Carnegie Mellon University, Stanford University, Columbia University, Texas A\&M University, University of Utah, University of Illinois--Urbana-Champaign, University of California--Los Angles, Cornell University, Rutgers University, University of Michigan, Northeastern University, University of Southern California, University of Minnesota, New York University, Purdue University, Princeton University, University of Wisconsin--Madison, University of Texas--Austin, University of Washington, and University of Pennsylvania.

The 20 departments randomly selected from the middle tier of the CSRankings include: University of Connecticut, University of Iowa, Clemson University,  University of North Carolina--Charlotte, University of Arizona, Iowa-State-University, University of Massachusetts-Lowell,  Rochester Institute of Technology, Illinois Institute of Technology, Drexel University, University of South Florida, University of Colorado--Colorado Springs, University of Tennessee, University of Georgia, University of Kentucky,  University of New Hampshire, University of Maryland--Baltimore County, University of Houston, Virginia Commonwealth University, and College of William \& Mary.

Finally, the 20 departments randomly selected from the bottom tier of the CSRankings were: University of Michigan--Dearborn,
University of Oklahoma, Missouri University of Science and Technology, University of Missouri, Montana State University, North Dakota State University, University of Arkansas--Little Rock, University of Miami, Nova Southeastern University, Mississippi State University, University of Nebraska--Omaha, Boston College, Boise State University, Texas Tech University, Georgia State University, University of Tulsa, New Mexico Institute of Mining and Technology, University of Wyoming, University of Wisconsin--Milwaukee, and University of Nevada--Las Vegas.

\begin{figure}
 \centerline{\includegraphics[width=2.75in]{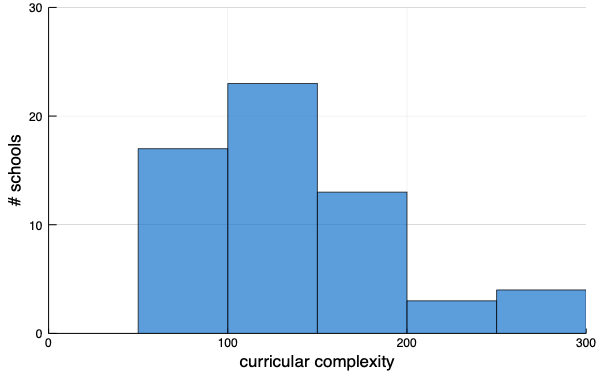}}
 \caption{The structural curricular complexity histogram for all curricula included in the study. The average complexity value of these curricula is $135.2$, with a standard deviation of $51.8$.}\label{all-tiers}
\end{figure}

\begin{figure*}
 \centerline{\includegraphics[width=\textwidth]{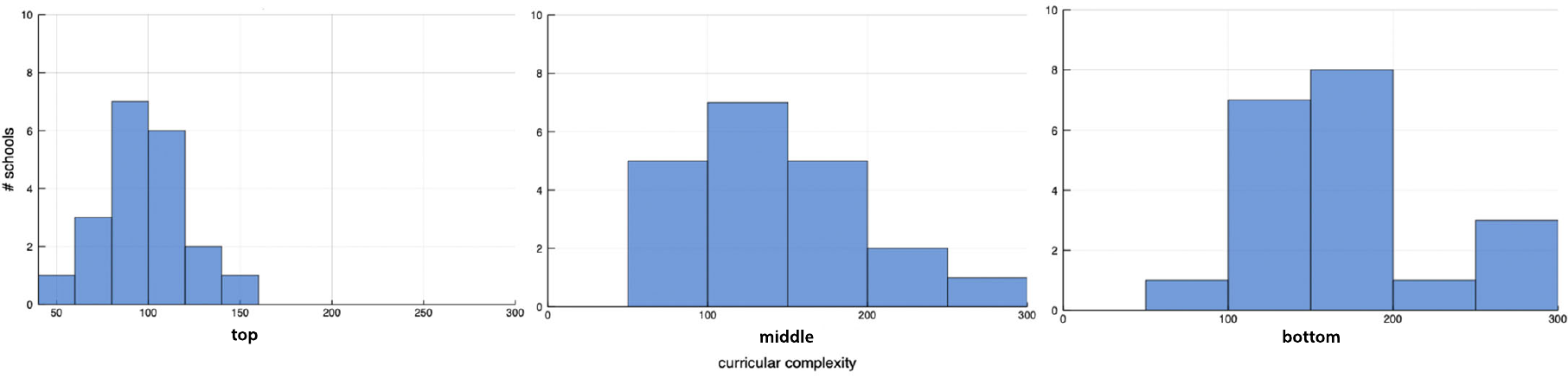}}
 \caption{The structural curricular complexity histograms of the departments in the study, disaggregated by tier. Note that these are approximately normal with similar variances. The top tier sample has an average curricular complexity of $96.7$ with a standard deviation of $21.6$. The mid tier sample has an average curricular complexity of $140.4$ with a standard deviation of $67.3$. The bottom tier sample has an average curricular complexity of $168.2$ with a standard deviation of $89.1$.}\label{tiers-disaggregated}
\end{figure*}

A histogram showing the structural curricular complexity distribution for all samples (i.e., the curricula sampled form all three tiers) is provided in Figure~\ref{all-tiers}.
The distribution of curricular complexities in this figure appears approximately Gaussian with $\mu = 135.2$ and $\sigma = 51.8$. However, when the curricular complexities are disaggregated according to the previously defined tiers, as shown in Figure~\ref{tiers-disaggregated}, possible differences appear. 

In order to test the null hypothesis using ANOVA, we must assume the structural curricular complexity values of the curricula within each tier are normally 
distributed, and that all three tiers have the same variance $\sigma^2$. It should be noted that these conditions can be moderately relaxed (particularly 
the normality assumption) and the analysis will remain valid~\cite{MeScWa:81}. 

The ANOVA method partitions the total sum of squares of the deviations in structural curricular complexity across all departments into two independent parts, one 
that is attributed to the independent variable (program quality in this case), and a remainder that is attributed to random errors arising from other 
factors not accounted for in this experiment. That is,
\begin{equation}
 TSS = SST + SSE,
  \label{TSS1}
\end{equation}
where $TSS$ denotes the total sum of squares of deviations, $SST$ represents the sum of squares of the deviations between the tiers, and SSE is 
the sum of squares attributed to errors or noise. More specifically, if we let $cc_{ij}$ denote the curricular complexity of the $j$-th curricula sampled 
from the $i$-th tier, then
\begin{equation}
 TSS = \sum_{i=1}^3 \sum_{j=1}^{n_i} \left(cc_{ij} - \overline{cc}\right)^2,
  \label{TSS2}
\end{equation}
where $\overline{cc}$ is the sample mean for all samples drawn over all tiers.  The sum of squares deviation between the tiers is given by 
\begin{equation}
 SST = \sum_{i=1}^3 n_i\left(\overline{T}_i - \overline{cc}\right)^2,
 \label{SST}
\end{equation}
where $T_i$ is the total structural curricular complexity of the sampled curricula from the $i$-th tier, and $\overline{T}_i = T_i/n_i, i = 1,2,3$, are the tier sample averages.
Note that when the sample means for the three tiers are the same, $SST=0$.

Substituting Equations~(\ref{TSS2}) and~(\ref{SST}) into Equation~(\ref{TSS1}) and solving for SSE yields:
\begin{equation}
 SSE = {\sum_{i=1}^3 \sum_{j=1}^{n_i} \left(cc_{ij} - \bar{T}_i\right)^2}.
\end{equation}

The unbiased estimator of $\sigma^2$ based on $n - 3$ degrees of freedom is given by the mean square error:
\begin{equation}
 MSE = {SSE \over n - 3},
 \label{MSE}
\end{equation}
where $n = n_1 + n_2 + n_3$.
The mean square for the tiers has 2 degrees of freedom, i.e., one less than the number of tiers, and is therefore
\begin{equation}
 MST = {SST \over 2}. 
 \label{MSE}
\end{equation}

In order to assess the statistical significance of a decision to reject the null hypothesis, an $F$-test is conducted to compare the deviation among the tier variances.  The $F$-test statistic is given by
\[
 F = {MST \over MSE}.
\] 
Note that the $F$-test is a ratio that compares the mean square variability between the tiers to the mean square variability within the tiers. Thus, 
as $F$-test values increase above $1$, the data are increasingly inconsistent with the null hypothesis, and the null hypothesis should be rejected 
when $F > F_{\alpha}$,
where $F_{\alpha}$ is the critical value of $F$ where the probability of a type I error is $\alpha$.

For the $F$ distribution with $(2, 54)$ degrees of freedom, $F_{0.05} = 3.15$.  That is, if the $F$-test for the experiment yields a value greater than $3.15$, we can reject the null hypothesis with only a 5\% chance of doing so in error.

Applying this ANOVA methodology to the aforementioned samples yielded the results shown in Table~\ref{ANOVA}. (This analysis is also available as a Jupyter notebook that uses the Curricular Analytics Toolbox~\cite{CA-toolbox}. To view this notebook, go to:  https://tinyurl.com/rbk2n9z.)
\begin{table}
\centering
\begin{tabular}{|l|c|c|c|c|}\hline
   ~        &      Sum of Squares       &  Deg. of Freedom	 &       Mean Square	     &    $F$ \\ \hline\hline
Tiers     & 46,735      &  2	 &  23,367   &   11.09 \\ \hline
Error     & 102,133      & 52 &   1,964     &              \\ \hline
Total     & 148,869	 & 54 &                  &              \\ \hline
\end{tabular}
\caption{The results of the ANOVA analysis associated with the samples selected from the three tiers of computer science departments. The  $F$-test statistic is $11.09$.}\label{ANOVA}
\end{table}
Notice that the $F$-test statistic obtained from this analysis is $11.09$.  Because
\[
  11.09 > F_{0.05} = 3.15,
\]
the null hypothesis should be rejected.  That is, with a low probability of error, the samples collected from each tier indicate that the mean curricular complexity values of the tiers are different.  This result, along with the evidence given in Figures~\ref{box-whisker} and~\ref{tiers-disaggregated}, provide strong evidence that higher quality computer science programs have lower structural curricular complexity, and that lower quality computer science programs have higher structural curricular complexity.  
 
\section{Discussion}
We have demonstrated that an inverse relationship exists between the structural complexity of the curricula in undergraduate computer science programs and 
the quality of these programs. Specifically, the complexity of the computer science undergraduate curricula at the highest quality programs (where quality 
is inferred from the rankings of computer science departments) is drastically less than the complexity of the curricula at those departments judged to be 
at the lower end of this quality ranking. The average curricular complexity of those departments at the bottom of the ranking is almost twice the average of those departments in the top decile of the ranking. In addition, we demonstrated that this difference is statistically significant; that is, this difference is due to 
something other than chance. 

There are reasonable causal arguments for both sides of the relationship we have demonstrated between structural curricular complexity and computer science program quality. First consider the complexity $\longrightarrow$ quality direction. Because the complexity of a curriculum is a measure of the difficulty that students are expected to have completing that curriculum, this difference has important student success implications. In particular, all other factors equal, we expect students to graduate at higher rates from lower complexity curricula. Graduation rates are an important consideration in rankings and the perceived quality of universities and their programs. An enhanced reputation allows a school to become more selective, and restrict admittance to better prepared students, which leads to higher graduation rates, which further enhances reputation, etc.  

One might further argue that because the top tier departments admit better prepared students they can offer less complex curricula, as their students can more easily overcome any knowledge gaps that may exist due to having fewer prerequisites prior to attempting a given course. It should be noted, however, that there are schools not considered highly selective that have created pathways in the first year of their STEM curricula that substantially reduce structural curricular complexity~\cite{HeHiSlAb:17}. These curricular innovations have been demonstrated to significantly improve graduation rates, as well as the attainment of program learning outcomes~\cite{KlBo:15}. That is, we believe it is possible to reduce the complexity of computer science programs that serve less-prepared students, while actually improving program quality (as judged by outcomes).
More generally, we have postulated that the principle of Occam's Razor, often applied to guide engineering designs towards the simplest and therefore best solutions, also applies to curriculum design. In other words, the simplest curriculum (in terms of structural complexity) that allows students to attain a program's learning outcomes yields the best student success outcomes.

A possible causal factor in the quality $\longrightarrow$ complexity direction involves the fact that higher quality programs tend to be much better resourced, as compared to lower quality programs. Thus, higher quality programs are able to apply more resources towards curricular matters~(e.g., faculty/staff time, investment in assessment and improvement, etc.), thus allowing these schools to provide more efficient curricula. Further studies may be constructed to determine the actual impact of these and possibly other causal factors. 

\begin{figure}
 \centerline{\includegraphics[width=6.5in]{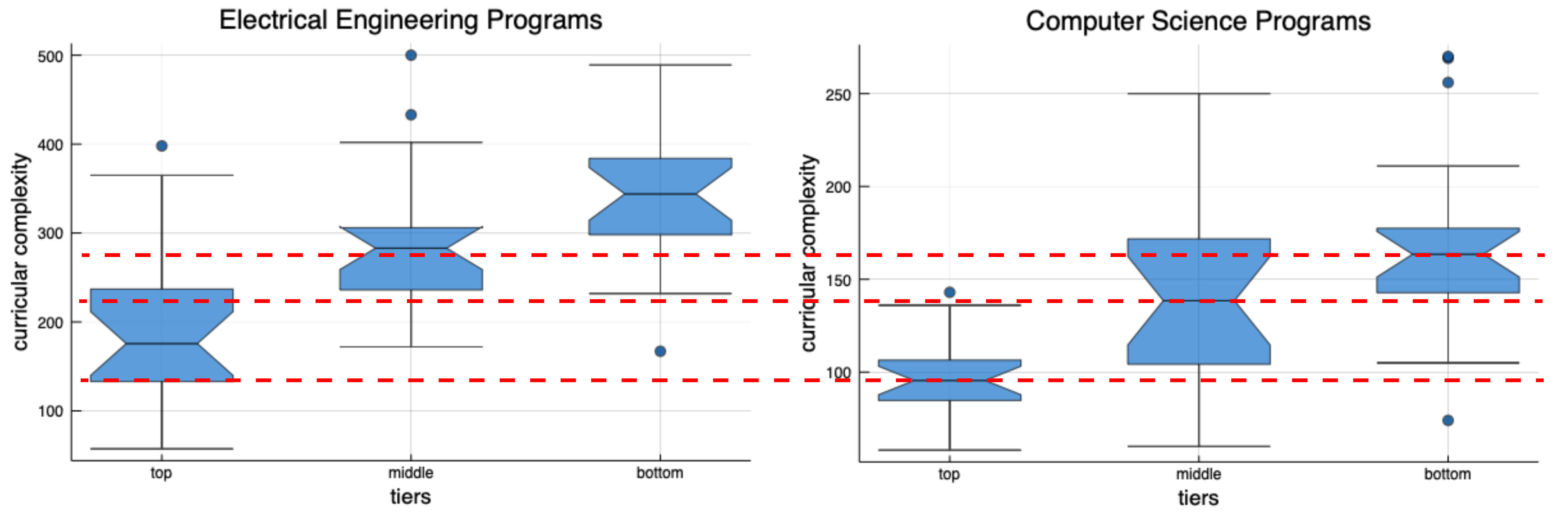}}
 \caption{A comparison of the structural complexities of undergraduate electrical engineering (left) and computer science (right) programs, broken down by quality tiers.  Dashed lines are drawn through the median complexity values for the top, middle, and bottom tier computer science programs. Notice that each is below the corresponding top, middle, and bottom tier electrical engineering programs.}\label{compare}
\end{figure}

Finally, it is interesting to consider how the structural complexity of high, medium, and bottom tier computer science program compare to high, medium, and bottom tier electrical engineering programs as shown in Figure~\ref{compare}~\cite{HeThAbFr:19}.
This suggests disciplinary differences may exist with respect to curricular complexity. Not only are computer science programs in general less structurally complex than electrical engineering programs, but this relationship is also preserved across tiers. The manner in which other engineering and STEM disciplines relate to this work warrants further investigation.

\bibliographystyle{abbrv} 
\bibliography{CS}


\end{document}